\documentclass[aps,prd,showpacs]{revtex4}%
\usepackage{amsmath}
\usepackage{graphicx}
\usepackage{epsfig}
\usepackage{amsfonts}
\usepackage{amssymb}%
\providecommand{\U}[1]{\protect\rule{.1in}{.1in}}
\begin{document}
\title{On the metric of the space of states in a modified QCD }
\author{Alejandro Cabo Montes de Oca}
\affiliation{Departamento de F\'{\i}sica Te\'{o}rica, Instituto de
Cibern\'{e}tica,Matem\'atica y F\'isica, Calle E, no. 309, Vedado, La Habana,
Cuba. }

\begin{abstract}
\noindent The form of the resulting Feynman propagators in a proposed local and gauge invariant QCD for massive fermions suggests the existence of indefinite metric associated to quark states, a property that
might relate it with the known Lee-Wick theories. Thus, the nature of the asymptotic free quark states in the theory is investigated here by quantizing the  quadratic part of the quark action. As opposite to
the case in the standard QCD, the free  theory does not show Hamiltonian constraints. The propagation modes include a family of massless waves and a complementary set of massive oscillations. The theory can be quantized in a way that the massive modes show positive metric and the massless ones exhibit negative norms. It is remarked that, since QCD is expected to not exhibit gluon or quark asymptotic states, the presence of negative metric massless modes does not constitute a definite drawback of the theory. In addition, the fact that the positive metric quark states are massive, seems to be a good feature of the model, being consistent with the approximate existence of asymptotically free massive states in high energy processes.

\end{abstract}

\maketitle

\section{Introduction}

In a recent work \cite{mQCD} it had been motivated an alternative for the
description of \ quarks interacting \ through \ gluons, which possible
equivalence with massless QCD was also argued. The outcome was a theory with
an action given by the massless QCD one, \ which also incorporates additional
terms, one for each quark flavour. The theory constitutes a local and gauge
invariant completion of the one discussed in Ref.
\cite{mpla,prd,epjc,epjc1,epjc2,epjc19,jhep,ana}, and, up to our knowledge, it
also furnishes a new local formulation for strongly interacting massive
quarks. The new Lagrangian terms appearing could also be interpreted as
possible counterterms, within \ a special renormalization scheme for massless
QCD. That is, it was pointed out that such terms could be present in an
effective action for massless QCD.\ These terms do not eliminate the power
counting renormalizability of the model, because the new quadratic part of the
action modifies the behavior of the free quark propagator, which now gets a
$1/p^{2}$ behavior at large momenta. The expansion is expected to be employed
in \ further studies directed to explore physical questions as: the
\ \ possibility of a flavour symmetry breaking effect determining a
hierarchical quark mass spectrum, and the properties of quark-quark
interaction potential in the scheme. However, the new diagrammatic expansion,
\ shows a \ conceptual feature which is convenient to be \ studied before
further employing it in the above cited studies. This special property is that
the new more convergent free propagator is given by the difference between a
free massless quark \ propagator \ and a massive one. This structure is very
similar to the typical ones appearing in the so called Lee-Wick theories, \ in
which usual \ bosons propagators are \ modified by substracting \ massive
propagators in order to make the expansion more \ convergent
\ \cite{lee-wick,lee-wick2,lee-wick3}. \ \ Such substractions, for the bosonic
fields, although making the propagators decreasing more rapidly at large
momenta, \ introduce states of negative norm in the space of states, which
mass shells are defined by the poles of the new propagators
\cite{nagy,nakanishi}. \ \ Thus, the quark modified propagator appearing in
reference \cite{mQCD}, \ which is the difference between \ a massless and a
massive one as described before, \ might in principle lead \ to the appearance
of \ an indefinite metric in the quark sector of the state space of the
theory. \ As remarked in Ref. \cite{lee-wick},\ the presence of an indefinite
metric could \ not directly imply a limitation in the physical applicability
of the theory. In our case this possibility might be indicated by the fact
that for QCD, it is conceived that there are not asymptotic states for the
elementary quarks and gluons. Also, the negative metric states in the Lee-Wick
theories can be compatible with an unitary scattering matrix by the
\ appearance of decaying rates for these negative norm states due to radiative
corrections. Thus, the presence of those modes on the proposed variation of
QCD might not automatically define its lack of applicability. \ \ However,
\ the existence of such modes, could \ \ occasionally lead \ to drastic
modifications of the predictions at \ very high energy processes, \ where the
\ perturbative expansion is expected to approximately work due to asymptotic
freedom. Thus, to know the type of \ negative metric states which can appear
in the \ proposed theory \ becomes an important question. It can be also noted
that the gluon free action is already including an indefinite metric
associated to the unphysical gluon components. \ However, since the gluon and
quark sectors do not couple in the free part of the action, the gluon free
theory quantization is identical as in the usual QCD.

\ Therefore, in this work, we investigate the \ quantization of the \ free
part of the quark action of the proposed theory. \ \ For this purpose we first
consider a reparametrization of the classical action obtained in \cite{mQCD}
in order to express the relativistic quark propagator in a form in which the
massive states appears as related with poles associated to \ positive metric
states and the massless ones on the contrary, as linked with negative metric
ones \cite{veltman}. Next, the description of the classical propagation modes
of the free Lagrangian equations is considered. Two sets of quark modes arise.
One of them is identical to the one corresponding to massless quarks in QCD,
and a second one is a set of massive states of polarizations being equivalent
to the usual set of modes satisfying \ the Dirac's equation. \ \ The classical
Hamiltonian \ and the canonical equations are also written. A helpful form of
the Hamiltonian, to be further employed for the quantization process is
obtained.\ It was used the specific form of the generalized Hamiltonian
procedure presented in Ref. \ \cite{nakano}.\ The consistency of the
Hamiltonian and Lagrangian equations were checked. \ \ Further \ the
\ quantization is considered by defining field operators as superpositions of
the previously determined wave modes. \ \ The Hamiltonian is then expressed in
terms of the creation and annihilation operators defining the massive and
massless quark fields. The result, as usual was written as a linear
combination of the number operators for each quark mode with the coefficient
defined by the energy for the mode. This conclusion followed after choosing
the form of the anticommutation relation among all the creation and
annihilation operators of the fermion modes, that define the quantum states as
showing positive norm for the massive quarks and negative ones for the
massless ones. Central in determining the chosen selection of the creation and
annihilation operators for the single particle states\ and the metric of the
state vector space, was, on one side, the requirement of defining a bounded
from below value of the energy of the system. The process to implementing
these properties was also basic in defining the \ positive or negative metric
character of the various quantized modes. The satisfaction of the same
classical field equations by the field operators and their momenta at all
times was shown. The quantum commutators between all the fields and their
momenta satisfied the usual quantization rule of being given by the imaginary
unit times the results of the classical Poisson brackets. \ \ \

\ The presentation proceeds as follows. \ In Section II, the general form of
the action for the theory is presented and the notations employed \ are
defined. Afterwards, the \ general action and the quark propagators are
rewritten, to fully evidence the similarity of the propagator structure with
the ones appearing in the Lee-Wick theories. This is done in a way suggesting
the positive metric of massive states and negative metric for massless states.
\ Section III, \ presents the determination of the quark modes associated to
the poles of the propagator. \ The Section IV considers the writing of the
Hamiltonian equations for the Grassman quark fields and their canonical
momenta. \ The equivalence of the canonical and Lagrangian equations is
verified and a formula for the Hamiltonian in terms of the fields is derived.
Finally, in Section V the fields are consistently quantized \ by properly
choosing anticommutation rules for the creation and annihilation operators for
all the modes, and selecting the positive metric for the massive quarks and a
negative metric for the massless states. \

\section{Modified QCD and the Lee-Wick theories}

\ The classical action defining the model introduced in Ref. \cite{mQCD} had
the explicit form%
\begin{align}
S  &  =\int dx\text{ \ }(-\frac{1}{4}F_{\mu\nu}^{a}(x)F^{a\mu\nu}(x)-\frac
{1}{2\alpha}\partial_{\mu}A^{a\mu}(x)\partial_{\nu}A^{a\nu}(x)+\overline
{c}^{a}(x)\partial_{\mu}D^{ab\mu}c^{b}(x)+\nonumber\\
&  \sum_{f}\overline{\Psi}_{f}^{i}(x)\text{ }i\gamma^{\mu}D_{\mu}^{ij}\Psi
_{f}^{j}(x)-\sum_{f}\varkappa\int dx\overline{\Psi}_{f}^{j}\text{ }%
(x)\gamma_{\mu}\overleftarrow{D}^{ji\mu}\text{ }\,\gamma_{\nu}D^{ik\nu}%
\Psi_{f}^{k}(x), \label{genaction}%
\end{align}
in which the conventions for the various quantities coincide with the ones
employed in Ref. \cite{muta}. \ In details, the appearing fields are defined
as follows \
\begin{align}
F_{\mu\nu}^{a}  &  =\partial_{\mu}A_{\nu}^{a}-\partial_{\nu}A_{\mu}%
^{a}-g\text{ }f^{abc}A_{\mu}^{b}A_{\nu}^{c},\nonumber\\
\Psi_{f}^{k}(x)  &  \equiv\left(
\begin{array}
[c]{c}%
\Psi_{f}^{k,1}(x)\\
\Psi_{f}^{k,2}(x)\\
\Psi_{f}^{k,3}(x)\\
\Psi_{f}^{k,4}(x)
\end{array}
\right)  ,\text{ }\\
\Psi_{f}^{\dag k}(x)  &  \equiv(\Psi_{f}^{k}(x))^{T\ast}=\left(
\begin{array}
[c]{cccc}%
(\Psi_{f}^{k,1}(x))^{\ast} & (\Psi_{f}^{k,2}(x))^{\ast} & (\Psi_{f}%
^{k,3}(x))^{\ast} & (\Psi_{f}^{k,4}(x))^{\ast}%
\end{array}
\right)  ,
\end{align}
where $f$ $=1,...,6$ indicates the flavour index. The expressions for the
Dirac conjugate spinors and covariant derivatives are
\begin{align}
\overline{\Psi}_{f}^{j}\text{ }(x)  &  =\Psi_{f}^{\dag k}(x)\gamma^{0},\text{
}\\
D_{\mu}^{ij}  &  =\partial_{\mu}\delta^{ij}-i\text{ }g\text{ }A_{\mu}^{a}%
T_{a}^{ij},\ \overleftarrow{D}_{\mu}^{ij}=-\overleftarrow{\partial}_{\mu
}\delta^{ij}-i\text{ }g\text{ }A_{\mu}^{a}T_{a}^{ij},\\
D_{\mu}^{ab}  &  =\partial_{\mu}\delta^{ab}-g\text{ }f^{abc}\text{ }A_{\mu
}^{c}.
\end{align}
in which the Dirac's matrices, $SU(3)$ generators and the metric tensor are
defined in the conventions of Ref. \cite{muta}, as
\begin{align}
\{\gamma^{\mu},\gamma^{\nu}\}  &  =2g^{\mu\nu},\ \ \ \ [T_{a},T_{b}]=i\text{
}f^{abc}T_{c}.\text{ }\gamma^{0}=\beta,\text{ }\gamma^{j}=\beta\text{ }%
\alpha^{j},\text{ }j=1,2,3,\nonumber\\
g^{\mu\nu}  &  \equiv\left(
\begin{array}
[c]{cccc}%
1 & 0 & 0 & 0\\
0 & -1 & 0 & 0\\
0 & 0 & -1 & 0\\
0 & 0 & 0 & 1
\end{array}
\right)  ,\text{ }\beta=\left(
\begin{array}
[c]{cc}%
I & 0\\
0 & -I
\end{array}
\right)  ,\text{ }\alpha^{j}\equiv\left(
\begin{array}
[c]{cc}%
0 & \sigma^{j}\\
\sigma^{j} & 0
\end{array}
\right)  ,\text{ }j=1,2,3,\nonumber\\
\sigma^{1}  &  =\left(
\begin{array}
[c]{cc}%
0 & 1\\
1 & 0
\end{array}
\right)  ,\text{ }\sigma^{2}=\left(
\begin{array}
[c]{cc}%
0 & -i\\
i & -I
\end{array}
\right)  ,\text{ }\sigma^{3}=\left(
\begin{array}
[c]{cc}%
1 & 0\\
0 & -1
\end{array}
\right)  ,\text{ }I=\left(
\begin{array}
[c]{cc}%
1 & 0\\
0 & 1
\end{array}
\right)  .
\end{align}
\

Other useful definitions and relations for the coordinates are
\[
x\equiv x^{\mu}=(x^{0},\overrightarrow{x})=(x^{0},x^{1},x^{2},x^{3}),\text{
\ }x_{\mu}=g_{\mu\nu}x^{\nu},\text{ }x^{0}=t.
\]

The complex conjugate operation is symbolized by the superscript $^{\ast}$,
\ by also assuming that the conjugate operation of a quantity also implies the
Hermitian conjugation of any operator entering its definition. For example, in
the conjugate transposed fermion field $\Psi_{f}^{\dag k}(x)=(\Psi_{f}%
^{k}(x))^{T\ast}$ the operator structure entering the definition of $\Psi
_{f}^{k}(x)$ should be substituted by its Hermitian transposed structure. In
what follows the color and spinor indices of the fields and other magnitudes
will be omitted in order to simplify the writing. The time variable $t$ will
be defined by $t=x^{0}.$ The system will be assumed to be enclosed in a large
cubic spatial box of volume $V$, \ on which periodic boundary conditions are
imposed for all the fields. This define the spatial momenta as taken \ the
usual discrete values.

\ The quadratic in the quark field of flavour $f$, free action of the theory
is given by
\begin{align}
\mathcal{S}_{0,f}  &  =\int dx\text{ }\overline{\Psi}_{f}(x)(i\gamma^{\mu
}\partial_{\mu}-\varkappa\text{ }\partial^{2})\Psi_{f}(x)\nonumber\\
&  =-\int dx\text{ }\overline{\Psi}_{f}\text{ }(x)\Lambda_{f}(\partial
)\Psi_{f^{\prime}}^{j}(x) \label{freeaction}%
\end{align}
in which \ $\varkappa$ is a constant with dimension of length, which the
discussion in former \ works motivating the proposal of the action
(\ref{genaction}) in Ref. \cite{mQCD}, suggested to be related with quark
condensation effects. \ Those analysis were associated to the \ consideration
of a fermion squeezed state as the vacuum leading to modified non local
perturbative Wick expansion for massless QCD . They strongly suggested the
local action form (\ref{genaction}) as possible effective action of the
massless QCD \cite{mpla,prd,epjc,epjc1,epjc2,epjc19,jhep,ana}.

The Feynman propagator \ $S_{f}$ for the considered quark of flavour $f$, \ is
given by the above defined inverse of the kernel $\Lambda_{f}$ \ and
satisfies
\begin{equation}
-(i\gamma^{\mu}\partial_{\mu}-\varkappa\text{ }\partial^{2})\text{ }%
S_{f}(x-y)=\delta(x-y),
\end{equation}
or in terms of its Fourier transform $\ S_{f}(x)=\int dp \,S_{f}%
(p)\exp(-ip.x)),$%
\begin{equation}
\text{ \ \ }-(\gamma^{\mu}p_{\mu}+\varkappa\text{ }p^{2})S_{f}(p)=I.
\end{equation}
In order to simplify the notation, in what follows we will use the same symbol
$I$ (or simply the unity $1$) for the Kronecker delta in the color and spinor
indexes, for both cases: the two component and the four component spinors,
when no confusion can arise.

Thus, \ the quark propagator for the quark of flavour $f$\ \ has the form
\begin{align}
S_{f}(p)  &  =\frac{1}{-\gamma_{\nu}p^{\nu}-\varkappa\text{ }p^{2}}\equiv
\frac{(-\gamma_{\nu}p^{\nu}-\lambda\text{ }p^{2})^{rr^{\prime}}\delta
^{ii^{\prime}}}{p^{2}(1-{\lambda}^{2}p^{2})}\nonumber\\
&  =-\frac{1}{\gamma_{\nu}p^{\nu}}-(-\frac{1}{\gamma_{\nu}p^{\nu}+m_{f}%
}),\text{ \ \ \ \ \ }m_{f}=\frac{1}{\varkappa}\text{. } \label{subst}%
\end{align}
The free quark propagator shows poles of mass $m_{f}=\frac{1}{\varkappa},$
\ which are inversely proportional to the parameter $\varkappa$. \ It also
present poles\ of zero mass. This propagator has the structure given in \ the
last line of (\ref{subst}), which expresses it, in the here employed notation,
as the usual massless quark propagator \ minus the Dirac propagator for \ a
massive quark in which the sign of the mass had been reversed. \ \ As
mentioned in the introduction, this structure closely resembles the \ Pauli
Villars kind of regularization employed in the Lee-Wick theories, which might
lead to the appearance of negative metric states in the resulting quantum
field theories. \ The massless component of the propagator shows the standard
form for massless QCD with positive metric, and the massive component has a
changed sign for the usual Dirac component of the action (as evidenced by the
large momentum limit of this term). This suggests that the quantization of the
Lagrangian \ with the employed field parametrization, will lead to negative
metric for the massive fermion sector of the asymptotic states. \ However,
since the theory shows \ space time inversion invariance, the sign of the time
which could effectively correspond with the physical time is not well defined
at the current stage of the analysis of the theory. Then, this freedom can be
considered as factor to employed in the process of constructing the theory.
After adopting this point of view, it seems possible to redefine the fields in
the action by performing the following change of variables and coordinates
within the functional integral defining the Green's functions generating
functional
\begin{align*}
x  &  =-x^{\prime},\text{ \ \ \ \ \ }\partial_{\mu}\equiv\frac{\partial
}{\partial\text{ }x^{\mu}}=-\partial^{\prime\mu}\equiv\frac{\partial}%
{\partial\text{ }x^{\prime\mu}},\\
\Psi(x)  &  =\Psi_{f}^{k}(-x^{\prime})=\Psi_{f}^{\prime k}(x^{\prime}),\text{
\ \ \ \ \ \ }\overline{\Psi}(x)=\overline{\Psi}_{f}^{k}(-x^{\prime}%
)=\overline{\Psi}_{f}^{\prime k}(x^{\prime}),\\
A^{a\mu}(x)  &  =A^{a\mu}(-x^{\prime})=-A^{\prime a\mu}(x^{\prime}),\text{ }\\
c(x)  &  =c(-x^{\prime})=c^{\prime}(x^{\prime}),\text{ \ }\overline
{c}(x)=\overline{c}(-x^{\prime})=\overline{c}^{\prime}(x^{\prime}).
\end{align*}

After performing this changes in the general expression for the action
(\ref{genaction}), its expression in terms of the new primed fields and
coordinates, for afterwards returning to the previous notation for the
coordinates and fields by removing the primes in all the magnitudes, it
follows for $S$%
\begin{align}
S  &  =\int dx\text{ }{\Large (}-\frac{1}{4}F_{\mu\nu}^{a}(x)F^{a\mu\nu
}(x)-\frac{1}{2\alpha}\partial_{\mu}A^{a\mu}(x)\partial_{\nu}A^{a\nu
}(x)+\overline{c}^{a}(x)\partial_{\mu}D^{ab\mu}c^{b}(x)-\nonumber\\
&  -\sum_{f}\overline{\Psi}_{f}^{i}(x)\text{ }i\gamma^{\mu}D_{\mu}^{ij}%
\Psi_{f}^{j}(x)-\sum_{f}\varkappa\int dx\overline{\Psi}_{f}^{j}\text{
}(x)\gamma_{\mu}\overleftarrow{D}^{ji\mu}\text{ }\gamma_{\nu}D^{ik\nu}\Psi
_{f}^{k}(x){\LARGE )}.\text{\ }%
\end{align}
\ The new expression for the free quark action takes the form%
\begin{equation}
\mathcal{S}_{0,f}=\int dx\text{ }\overline{\Psi}_{f}(x)(-i\gamma^{\mu}%
\partial_{\mu}-\varkappa\text{ }\partial^{2})\Psi_{f}(x)=-\int dx\text{
}\overline{\Psi}_{f}\text{ }(x)\Lambda_{f}(\partial)\Psi_{f^{\prime}}^{j}(x),
\label{freeaction2}%
\end{equation}
which defines for the new propagator kernel
\begin{align}
S_{f}(p)  &  =\frac{1}{\gamma_{\nu}p^{\nu}-\varkappa\text{ }p^{2}}\nonumber\\
&  =\frac{1}{\gamma_{\nu}p^{\nu}}-(\frac{1}{\gamma_{\nu}p^{\nu}-m_{f}}),
\end{align}
where now the massive term \ has the form of the \ typical Feynman quark
propagator for massive quarks and the massless one has a different sign, which
suggests its association to negative metric states. In order to define the
precise way in which the theory \ implements \ the introduction of a space of
states with indefinite metric for quarks, in next sections we perform the
explicit quantization of this new form of the action.

\section{Classical quark propagation modes}

Let us \ determine in this section the \ quark propagation modes defined by
the free fermion component of the action. These $u(x)$ modes satisfies the
\ Lagrange equations defined by (\ref{freeaction2}), which are
\begin{equation}
(-i\gamma^{\mu}\partial_{\mu}-\varkappa\text{ }\partial^{2})\text{
}u(x)=0,\text{ \ }u(x)=\int dq^{4}u(q)\exp(-q_{\mu}x^{\mu}). \label{Free}%
\end{equation}
\ They imply for the Fourier components $u(q),$ the matrix equation%
\begin{equation}
(-\gamma^{\mu}q_{\mu}+\varkappa\text{ }q^{2})\text{ }u(q)=0. \label{Lmassive}%
\end{equation}
This expression clearly shows that there exist massless as well as massive
propagation modes. Let us separately consider in what follows the massive and
massless solutions.

\subsection{Massive solutions}

\ \ The Fourier four momentum $q$ associated to \ any of the waves being
determined, will be expressed in the form $q=(q^{0},q^{i}).$ \ Since we are
searching for the massive modes, in the Lorentz frame in which the momentum
vanishes, the momentum can be written in the form $\ \ q=(q^{0},0),$ and thus,
the equation (\ref{Lmassive}) \ reduces to%
\begin{equation}
q_{0}(-\gamma^{0}+xq_{0})\,u(q)=0\equiv q_{0}\left(
\begin{array}
[c]{cc}%
(-1+\varkappa\text{ }q_{0})I & 0\\
0 & (\varkappa\text{ }q_{0}+1)I
\end{array}
\right)  \left(
\begin{array}
[c]{c}%
u^{(1)}\\
u^{(2)}%
\end{array}
\right)  , \label{equation}%
\end{equation}
which clearly allows to easily define in this frame the solutions for the
separate cases in which \ the energy component $q_{0}$ takes positive or
negative values. \ Both cases are considered in what follows in this subsection.

\subsubsection{Positive energy waves}

Below, it will be convenient to define a mass parameter $m_{f}$ as
\begin{equation}
\varkappa=\frac{1}{m_{f}}.
\end{equation}
For this case of positive energy waves, in the selected rest frame, the energy
$q_{0},$ takes the positive value $q_{0}=m_{f}$ and the positive energy spinor
polarization gets the simple form
\begin{align*}
u^{r}(\overrightarrow{0})  &  =\left(
\begin{array}
[c]{c}%
\beta^{r}\\
0
\end{array}
\right)  , r=1,2\\
\beta^{1}  &  =\left(
\begin{array}
[c]{c}%
1\\
0
\end{array}
\right)  ,\beta^{2}=\left(
\begin{array}
[c]{c}%
0\\
1
\end{array}
\right)  .
\end{align*}
Then, the massive positive energy polarization of the considered theory
coincide with the ones associated to the positive energy modes of the usual
solutions of the Dirac's equation for a massive particle. Therefore, the
spinor polarizations in any Lorentz frame might be written in the form
\begin{align}
u^{r}(\overrightarrow{p})  &  =\sqrt{\frac{\epsilon_{m}(\overrightarrow{p}%
)+m_{f}}{2\epsilon_{m}(\overrightarrow{p})}}\left(
\begin{array}
[c]{c}%
\beta^{r}\\
\frac{\overrightarrow{\sigma}\cdot\overrightarrow{p}}{\epsilon_{m}%
(\overrightarrow{p})+m_{f}}\beta^{r}%
\end{array}
\right)  ,r=1,2,\label{massive1}\\
\overrightarrow{\sigma}.\overrightarrow{p}  &  =\sigma^{i}p^{i},\text{
\ }\epsilon_{m_{f}}(\overrightarrow{p})=\sqrt{m_{f}^{2}+\overrightarrow{p}%
^{2}},q=(\epsilon_{m_{f}}(\overrightarrow{p}),\overrightarrow{p}),\nonumber
\end{align}
in which the energies of the modes are given by $\epsilon_{m_{f}%
}(\overrightarrow{p})=\sqrt{m_{f}^{2}+\overrightarrow{p}^{2}}.$ The
polarizations had been chosen to satisfy the normalization conditions%
\[
u^{r\dag}(\overrightarrow{p})u^{s}(\overrightarrow{p})=\delta^{rs}%
,\ \ \overline{u}^{r}(\overrightarrow{p})u^{s}(\overrightarrow{p})=\frac
{m_{f}}{\epsilon_{m_{f}}(\overrightarrow{p})}\delta^{rs}.
\]

Note that the normalization for the spinors is fixed in a slightly different
way as the usual one. This was done in order to assure the normalization to
one of the spinors in \ the large quantization volume for the usual spinor
scalar product.

\subsubsection{Negative energy waves}

In a similar way, for the negative energy case $\ q_{0}=-m_{f}$ \ the
solutions of the equation (\ref{equation}) leads to the spinor polarization of
the form%
\begin{align}
v^{r}(\overrightarrow{p})  &  =\sqrt{\frac{\epsilon_{m}(\overrightarrow{p}%
)+m_{f}}{2\epsilon_{m}(\overrightarrow{p})}}\left(
\begin{array}
[c]{c}%
\frac{-\overrightarrow{\sigma}\cdot\overrightarrow{p}}{\epsilon_{m}%
(\overrightarrow{p})+m_{f}}\beta^{r}\\
\beta^{r}%
\end{array}
\right)  ,r=1,2,\label{massive2}\\
\overrightarrow{\sigma}.\overrightarrow{p}  &  =\sigma^{i}p^{i},\text{
\ }\epsilon_{m_{f}}(\overrightarrow{p})=\sqrt{m_{f}^{2}+\overrightarrow{p}%
^{2}},q=(-\epsilon_{m_{f}}(\overrightarrow{p}),\overrightarrow{p}),\nonumber
\end{align}
in which again the energy parameter $\epsilon_{m_{f}}(\overrightarrow{p}%
)=\sqrt{m_{f}^{2}+\overrightarrow{p}^{2}}$ and the polarizations were chosen
to satisfy the normalization conditions%
\[
v^{r+}(\overrightarrow{p})\text{ }v^{s}(\overrightarrow{p})=\delta^{rs},\text{
\ \ }\overline{v}^{r}(\overrightarrow{p})\text{ }v^{s}(\overrightarrow{p}%
)=-\frac{m_{f}}{\epsilon_{m_{f}}(\overrightarrow{p})}\delta^{rs}.
\]
The four momenta of the negative energy waves is given by $\ q=(-\epsilon
_{m_{f}}(\overrightarrow{p}),\overrightarrow{p}).$ Note that as usual, in
order to assure orthonormality for different values of the spatial momentum
$\overrightarrow{p}$ \ of the modes, the spatial momentum index in the Fourier
expansion of the original wave $u(x),$ had been changed in sign for the
negative energy solutions.

\subsection{Massless solutions}

These solutions of \ the general equation (\ref{Lmassive}), since having
\ $q^{2}=(q^{0})^{2}-\overrightarrow{q}^{2}=0$ , \ satisfy the \ massless
Dirac equation%
\begin{equation}
i\text{ }\gamma^{\mu}\partial_{\mu}\text{ }u(x)=0,\text{ }u(x)=\int dq\text{
}u(q)\exp(-q_{\mu}x^{\mu}),
\end{equation}
which for the Fourier components $u(q)$ take the form%
\begin{equation}
\gamma^{\mu}q_{\mu}\text{ }u(q)=0.
\end{equation}
\ In terms of the null four momentum $\ q=(q^{0},q^{i})$ it can be written as
follows%
\begin{align*}
(q_{0}\gamma^{0}-\gamma^{i}q^{i})u(q)  &  =\beta(q_{0}-\alpha^{i}q^{i})u(q)\\
&  \equiv\beta\left(
\begin{array}
[c]{cc}%
q_{0}I & -\overrightarrow{\sigma}.\overrightarrow{q}\\
-\overrightarrow{\sigma}.\overrightarrow{q} & q_{0}I
\end{array}
\right)  \left(
\begin{array}
[c]{c}%
\beta_{1}\\
\beta_{2}%
\end{array}
\right)  =0.
\end{align*}

It is clear that by considering the two component spinors $\beta_{1}$ and
$\beta_{2}$ as eigenfunctions of the helicity operator $\overrightarrow{\sigma
}.\overrightarrow{q}$, the positive and negative energy polarizations \ can be
determined. Below we examine the two cases.

\subsubsection{Positive energy massless waves}

\ After defining the $\beta^{+1}(\overrightarrow{p}),\beta^{-1}%
(\overrightarrow{p})$ as the eigenfunctions of $\overrightarrow{\sigma
}.\overrightarrow{p},$ given by \cite{schweber}%
\begin{align}
\beta^{+1}(\overrightarrow{p})  &  \equiv\frac{1}{\sqrt{2(n_{3}+1)}}\left(
\begin{array}
[c]{c}%
n_{3}+1\\
n_{1}+i\text{ }n_{2}%
\end{array}
\right)  ,\beta^{-1}(\overrightarrow{p})=\frac{1}{\sqrt{2(n_{3}+1)}}\left(
\begin{array}
[c]{c}%
-n_{1}+i\text{ }n_{2}\\
n_{3}+1
\end{array}
\right)  ,\\
\text{\ }\overrightarrow{\sigma}.\overrightarrow{p}\text{ }\beta
^{l}(\overrightarrow{p})  &  =l\text{ }\beta^{l}(\overrightarrow{p}),\text{
\ \ }l=+1,-1,\nonumber\\
\overrightarrow{n}(\overrightarrow{p})  &  =\frac{\overrightarrow{p}%
}{|\overrightarrow{p}|}=(n_{1},n_{2},n_{3}),\nonumber
\end{align}
the positive energy solutions \ of \ momentum $q=(\epsilon_{0}%
(\overrightarrow{p}),\overrightarrow{p})$ can be written in the form%
\begin{align}
u_{0}^{l}(\overrightarrow{p})  &  =\sqrt{\frac{1}{2}}\left(
\begin{array}
[c]{c}%
\beta^{l}(\overrightarrow{p})\\
l\text{ }\beta^{l}(\overrightarrow{p})
\end{array}
\right)  ,\text{ }l=+1,-1,\label{massless1}\\
\text{ \ }\epsilon_{0}(\overrightarrow{p})  &  =|\overrightarrow{p}|,\text{
\ \ \ \ \ }q=(\epsilon_{0}(\overrightarrow{p}),\overrightarrow{p}).\nonumber
\end{align}

The normalization of these modes satisfy%
\begin{equation}
u_{0}^{\dag\text{ }l}(\overrightarrow{p})\text{ }u_{0}^{l^{\prime}%
}(\overrightarrow{p})=\text{\ }\delta^{l\text{ }l^{\prime}},\text{
\ }\overline{u}_{0}^{l}(\overrightarrow{p})\text{ }u_{0}^{l^{\prime}%
}(\overrightarrow{p})=0.
\end{equation}

\subsubsection{Negative energy massless waves}

The negative energy modes, defined by $q_{0}=-|\overrightarrow{q}|$ , \ have
the polarization%
\begin{align}
v_{0}^{l}(\overrightarrow{p})  &  =\sqrt{\frac{1}{2}}\left(
\begin{array}
[c]{c}%
\beta^{-l}(\overrightarrow{p})\\
l\text{ }\beta^{-l}(\overrightarrow{p})
\end{array}
\right)  ,l=+1,-1,\label{massless2}\\
\overrightarrow{n}(\overrightarrow{p})  &  =\frac{\overrightarrow{p}%
}{|\overrightarrow{p}|}=(n_{1},n_{2},n_{3}),\nonumber
\end{align}
where, as it is convenient for assuring orthonormality of the modes for
different spatial momentum index, the spatial Fourier momentum
$\overrightarrow{q}$ was fixed to be equal to \ minus $-\overrightarrow{p}.$
\ The appearing two component spinors have the same already defined explicit
expressions%
\begin{equation}
\beta^{+}(\overrightarrow{p})\equiv\frac{1}{\sqrt{2(1+n_{3})}}\left(
\begin{array}
[c]{c}%
1+n_{3}\\
n_{1}+i\text{ }n_{2}%
\end{array}
\right)  ,\beta^{-}(p)=\frac{1}{\sqrt{2(1+n_{3})}}\left(
\begin{array}
[c]{c}%
-n_{1}+i\text{ }n_{2}\\
1+n_{3}%
\end{array}
\right)  ,
\end{equation}
and the four Fourier momentum of the mode has the form \ $q=(-\epsilon
_{0}(\overrightarrow{p}),\overrightarrow{p})$ with $\ \ \epsilon
_{0}(\overrightarrow{p})=|\overrightarrow{p}|.$ These polarizations obey the
orthonormality relations%
\begin{equation}
v_{0}^{l\dag}(\overrightarrow{p})\text{ }v_{0}^{l^{\prime}}(\overrightarrow{p}%
)=\text{\ }\delta^{l\text{ }l^{\prime}},\text{ \ }\overline{v}_{0}%
^{l}(\overrightarrow{p})\text{ }v_{0}^{l^{\prime}}(\overrightarrow{p})=0.
\end{equation}

\section{Free quark field quantization}

In this section we firstly present the application of the generalized
Hamiltonian formalism to the \ free part of the quark action of the modified
QCD in the form (\ref{freeaction2}). This discussion will also permit \ to
obtain a formula for the Hamiltonian of the system which is further employed
\ to quantize the theory.

\subsection{Generalized canonical procedure}

\ After performing few integrations \ by parts, the classical action $S$ in
(\ref{freeaction2}) can be expressed as a functional of the \ quark fields and
their first time derivatives in a form which determines for the Lagrangian $L$
of the system
\begin{align}
L  &  =\int dx^{3}\Psi^{\dag}(x){\huge (}-\frac{i}{2}\alpha^{\mu
}\overleftrightarrow{\partial}_{\mu}+\varkappa\text{ }(\overleftarrow{\partial
}_{t}\text{ }\overrightarrow{\partial}_{t}-\overleftarrow{\nabla
}.\overrightarrow{\nabla})\beta{\huge )}\Psi(x),\label{L}\\
\overleftrightarrow{\partial}_{\mu}  &  =\overrightarrow{\partial}_{\mu
}-\overleftarrow{\partial}_{\mu},\alpha^{\mu}=(I,\alpha^{1},\alpha^{2}%
,\alpha^{3}),\Psi^{\dagger}(x)=\Psi^{\ast T}(x),
\end{align}
where the right or left arrows appearing over the derivatives indicate whether
they are acting on the right or the left expressions respectively. It can be
helpful to recall that when operators will appear, the complex conjugation
operation, designed by the superindex "$^{\ast}$" also includes the Hermitian
transpose of the operator structure.

\ The momenta associated to the, up to now, classical fields $\Psi$ and
$\Psi^{\dagger}$ are defined in the generalized mechanics with anticommuting
variables (See Ref. \cite{nakano})\ in the form%
\begin{align}
\Pi_{\Psi}(x)  &  =L\frac{\overleftarrow{\delta}}{\delta(\partial_{t}\Psi
(x))}=-\frac{i}{2}\Psi^{\dagger}(x)+\varkappa\text{ }\partial_{t}\Psi
^{\dagger}(x)\beta,\\
\Pi_{\Psi^{\dagger}}(x)  &  =L\frac{\overleftarrow{\delta}}{\delta
(\partial_{t}\Psi^{\dagger}(x))}=-\frac{i}{2}\Psi(x)-\varkappa\text{ }%
\beta\text{ }\partial_{t}\Psi(x),
\end{align}
in which the "right" \ Grassman functional derivative employed is defined for
functionals which depend of the field and their derivatives as functions of
the spatial variables $\overrightarrow{x}$ at a time slice defined by a
instant $t$. This instant is given by the fixed time slice employed to perform
the spatial volume integration \ defining the Lagrangian in \ (\ref{L}).
\ \ It is interesting that this fermion theory is regular, since it does not
have \ Hamiltonian constraints. Then, all the velocities can be \ explicitly
expressed \ in terms of the \ coordinates (fields) and their momenta as
follows%
\begin{align}
\partial_{t}\Psi^{\dagger}(x)\beta &  =\frac{1}{\varkappa}(\frac{i}{2}%
\Psi^{\dagger}(x)+\text{ }\Pi_{\Psi}(x)),\\
\beta\partial_{t}\Psi(x)  &  =-\frac{1}{\varkappa}(\Pi_{\Psi^{\dagger}%
}(x)+\frac{i}{2}\Psi(x)).
\end{align}

With these expressions, the Hamiltonian at the defined time slice can be
expressed as a functional of the \ fields and momenta in the way
\begin{align}
H  &  =\int dx^{3}(\Pi_{\Psi}(x)\partial_{t}\Psi(x)+\Pi_{\Psi^{\dagger}%
}(x)\partial_{t}\Psi^{\dagger}(x))-L\nonumber\\
&  =\int dx^{3}{\huge \{}\frac{1}{\varkappa}{\Large (}-\Pi_{\Psi}(x)\text{
}\beta\text{\ }\Pi_{\Psi^{\dagger}}(x)+\frac{1}{4}\Psi^{\dagger}(x)\text{
}\beta\text{ }\Psi(x)-\nonumber\\
&  \frac{i}{2}\Psi^{\dagger}(x)\text{ }\beta\text{ }\Pi_{\Psi^{\dagger}%
}(x)-\frac{i}{2}\Pi_{\Psi}(x)\text{ }\beta\text{ }\Psi(x){\Large )}%
+\nonumber\\
&  \text{\ }\frac{i}{2}\Psi^{\dagger}(x)\alpha^{i}\overleftrightarrow{\partial
}_{i}\Psi(x)+\varkappa\text{ }\Psi^{\dagger}(x)\overleftarrow{\nabla
}.\overrightarrow{\nabla}\beta\Psi(x){\huge \}}. \label{ham}%
\end{align}

The expression for the Hamiltonian, allows to write \ the canonical equations
of motion in the form \cite{nakano}%
\begin{align}
\partial_{t}\Psi(x)  &  =\frac{\overrightarrow{\delta}}{\delta\Pi_{\Psi}%
(x)}H=-\frac{1}{\varkappa}\beta(\Pi_{\Psi^{\dagger}}(x)+\frac{i}{2}%
\Psi(x)),\nonumber\\
\partial_{t}\Psi^{\dagger}(x)  &  =\frac{\overrightarrow{\delta}}{\delta
\Pi_{\Psi^{\dagger}}(x)}H=\frac{1}{\varkappa}(\Pi_{\Psi}(x)+\frac{i}{2}%
\Psi^{\dagger}(x))\beta\nonumber\\
\partial_{t}\Pi_{\Psi}(x)\text{ }  &  =-H\frac{\overleftarrow{\delta}}%
{\delta\Psi(x)}=\frac{i}{2\varkappa}\Pi_{\Psi}(x)\beta-\frac{1}{4\varkappa
}\Psi^{\dagger}(x)\beta\nonumber\\
&  +i\Psi^{\dagger}(x)\alpha^{i}\overleftarrow{\partial}_{i}+\varkappa
\nabla^{2}\Psi^{\dagger}(x)\beta,\nonumber\\
\partial_{t}\Pi_{\Psi^{\dagger}}(x)  &  =-H\frac{\overleftarrow{\delta}%
}{\delta\Psi^{\dagger}(x)}=-\frac{i}{2\varkappa}\beta\text{ }\Pi
_{\Psi^{\dagger}}(x)+\frac{1}{4\varkappa}\beta\text{ }\Psi(x)\nonumber\\
&  +i\text{ }\alpha^{i}\partial_{i}\Psi(x)-\varkappa\beta\nabla^{2}\Psi(x),
\label{caneq}%
\end{align}
which after eliminating the momenta in terms of the fields and their time
derivatives lead to the Lagrange equations%
\[
(-i\text{ }\gamma^{\mu}\partial_{\mu}-\varkappa\text{ }\partial^{2}%
)\Psi(x)=0,
\]
in consistency with the starting ones in (\ref{Free}).

\subsection{Quantization}

Let us consider in this section the quantization of the quark fields $\Psi$
and $\Psi^{\dagger}.$ For this purpose it seems convenient to find a formula
\ for the Hamiltonian $H$ \ in terms of the fields and their time derivatives.
\ In the case of the Dirac's equation, and thanks to its constrained canonical
structure, the Hamiltonian can be \ expressed in terms of the \ fields $\Psi$
and $\Psi^{\dagger}$ and their spatial derivatives. However, the here
considered mechanical system is regular and \ the formula for the enenrgy
should also contain the time derivatives of the fields. \ \ The expression for
$H$ can \ be rewritten after eliminating the momenta in (\ref{ham}) by
employing \ the canonical equations (\ref{caneq})%
\begin{align}
H  &  =\int dx^{3}{\huge \{}\varkappa\text{ }\partial_{t}\Psi^{\dagger
}(x)\text{ }\beta\text{ }\partial_{t}\Psi(x)+i\text{ }\Psi^{\dagger}%
(x)\alpha^{i}\partial_{i}\Psi(x)+\varkappa\,\Psi^{\dagger}(x)\text{
}\overleftarrow{\nabla}.\overrightarrow{\nabla}\beta\Psi(x){\huge \}}%
\nonumber\\
&  =\int dx^{3}\Psi^{\dagger}(x){\huge \{}\varkappa\overleftarrow{\partial
}_{t}\text{ }\beta\text{ }\overrightarrow{\partial}_{t}+i\text{ }\alpha
^{i}\partial_{i}+\varkappa\text{ }\overleftarrow{\nabla}%
.\overrightarrow{\nabla}\beta{\huge \}}\Psi(x)\nonumber\\
&  =\int dx^{3}\Psi^{\dagger}(x)\,h(\partial)\,\Psi(x), \label{ham2}%
\end{align}
in which, as it was remarked before, the time derivatives of the fields enter
the definition of the \ kernel $h.$ \

\ Let us now start the quantization procedure \ by considering the quark
fields as \ superpositions of all the \ oscillator modes with coefficients
defined by \ operators. They will be identified as \ creation and annihilation
operators for all the modes. For this purpose it will be useful to separate
the field as a sum of two components: \ one associate to the \ massless modes
and another defined by the massive ones. The kind of the mode \ will be
represented by the greek index $\kappa$ taking two values: $0$ or $m.$ Then
the total field \ will be written as follows
\begin{align}
\Psi(x)  &  =\Psi_{m}(x)+\Psi_{0}(x)=\sum_{\varkappa=0,m}\Psi_{\kappa}(x),\\
\Psi_{0}(x)  &  =\frac{1}{\sqrt{V}}\sum_{\overrightarrow{p}}\sum_{l=\pm
1}{\Large (}a_{l}(\overrightarrow{p})u_{0}^{l}(\overrightarrow{p}%
)\exp(-i\text{ }(\epsilon_{0}(\overrightarrow{p})t-\overrightarrow{p}%
.\overrightarrow{x}))+\nonumber\\
&  \text{ \ \ \ \ \ \ \ \ \ \ \ \ \ \ \ \ \ \ \ \ \ \ \ \ \ }c_{l}^{\dag
}(\overrightarrow{p})v_{0}^{l}(\overrightarrow{p})\exp(i\text{ }(\epsilon
_{0}(\overrightarrow{p})t+\overrightarrow{p}.\overrightarrow{x})){\Large )}%
\nonumber\\
&  =u_{0}(x)+v_{0}(x)\\
\Psi_{m}(x)  &  =\frac{1}{\sqrt{V}}\sum_{\overrightarrow{p}}\sum
_{r=1,2}\text{\ }{\Large (}b_{r}(\overrightarrow{p})u_{m_{f}}^{r}%
(\overrightarrow{p})\exp(-i\text{ }(\epsilon_{m_{f}}(\overrightarrow{p}%
)t-\overrightarrow{p}.\overrightarrow{x}))+\nonumber\\
&  \text{\ \ \ \ \ \ \ \ \ \ \ \ \ \ \ \ \ \ \ \ \ \ \ \ \ }d_{r}^{\dag
}(\overrightarrow{p})v_{m_{f}}^{r}(\overrightarrow{p})\exp(i\text{ }%
(\epsilon_{m_{f}}(\overrightarrow{p})t+\overrightarrow{p}.\overrightarrow{x}%
){\Large )}\nonumber\\
&  =u_{m}(x)+v_{m}(x),
\end{align}
where the polarization vectors for all the massless or massive modes were
before determined. The assignation of the creation or annihilation nature for
each of the eight operators ($a_{l}(\overrightarrow{p}),$ $a^{\dag}
_{l}(\overrightarrow{p}),c_{l}(\overrightarrow{p}),c_{l}^{\dag}%
(\overrightarrow{p}),b_{r}(\overrightarrow{p}),b_{r}^{\dag}(\overrightarrow{p}%
),d_{r}(\overrightarrow{p}),d_{r}^{\dag}(\overrightarrow{p})$) is not yet
defined at this level. It will be fixed afterwards, by the requirements to be
imposed on the metric of the state space to define a bounded from below energy
for the many particle states. Thus, up to now \ the symbols $\dag$ over any of
these quantities will only mean that it is the Hermitian conjugate of the
operator, not that it is a creation one.

\ One important point to note is that \ the expression (\ref{ham2}) is defined
by the spatial integral in a given fixed time slice. Thus the time derivatives
of the fields entering, will be evaluated by using the time dependences of the
\ classical modes. \ \ Therefore, for the sake of consistency \ it should be
required that after quantization of this free theory, the expression of the
Hamiltonian $H$ should reproduce the classical time dependency of the fields
through the unitary transformation defining the operators in the Heisenberg
representation $\Psi(\overrightarrow{x},t)=\exp(iH$ $t)\Psi(\overrightarrow{x}%
,0)\exp(-iH$ $t).$

\ In addition, the field will be also decomposed in the sum of its the terms
associated to the positive energy modes $u_{\varkappa}(x)$, plus the terms
defined by the negative energy modes $v_{\varkappa}(x),$ for each \ of the two
values of the mass $\kappa$, as follows%
\begin{equation}
\Psi(x)=\sum_{\kappa=0,m}u_{\kappa}(x)+\sum_{\kappa=0,m}v_{\kappa}(x).
\end{equation}

In terms of this expansion the \ Hamiltonian has the expression%
\begin{align}
H  &  =\int dx^{3}\Psi^{\dagger}(x)\text{ }h(\partial)\text{ }\Psi
(x)\nonumber\\
&  =\sum_{\kappa=0,m}\sum_{\kappa^{\prime}=0,m}{\LARGE (}\int dx^{3}u_{\kappa
}^{\dagger}(x)h(\partial)u_{\kappa^{\prime}}(x)+\int dx^{3}v_{\kappa}%
^{\dagger}(x)h(\partial)v_{\kappa^{\prime}}(x)+\nonumber\\
&  \text{ \ \ \ \ \ \ \ \ \ \ \ \ \ \ \ \ \ \ \ \ \ \ \ \ }\int dx^{3}%
u_{\kappa}^{\dagger}(x)h(\partial)v_{\kappa^{\prime}}(x)+\int dx^{3}v_{\kappa
}^{\dagger}(x)h(\partial)u_{\kappa^{\prime}}(x){\LARGE )}.
\end{align}

The procedures for the calculation of the appearing matrix elements of the
\ kernel $h(\partial)$ are illustrated in \ the Appendix A. \ It follows that
they vanish when both entering modes\ have different \ $u$ or $v$ types \ and
also when \ they have different \ masses. \ \ The calculation of the few non
vanishing terms leads to the following formula for the \ Hamiltonian operator%
\begin{align}
H  &  =\int dx^{3}\Psi^{\dagger}(x)\text{ }h(\partial)\text{ }\Psi
(x)\nonumber\\
&  =\int dx^{3}u_{m}^{\dagger}(x)\text{ }h(\partial)\text{ }u_{m}(x)+\int
dx^{3}v_{m}^{\dagger}(x)\text{ }h(\partial)\text{ }v_{m}(x)+\nonumber\\
&  \int dx^{3}u_{0}^{\dagger}(x)\text{ }h(\partial)\text{ }u_{0}(x)+\int
dx^{3}v_{0}^{\dagger}(x)\text{ }h(\partial)\text{ }v_{0}(x)\nonumber\\
&  =\sum_{\overrightarrow{p}}\sum_{l=\pm1}{\large (-}\epsilon_{0}%
(\overrightarrow{p})\text{ }a_{l}^{\dag}(\overrightarrow{p})a_{l}%
(\overrightarrow{p})+\epsilon_{0}(\overrightarrow{p})\text{ }c_{l}%
(\overrightarrow{p})c_{l}^{\dag}(\overrightarrow{p}){\large )}+\nonumber\\
&  \sum_{\overrightarrow{p}}\sum_{r=1,2}{\large (}\epsilon_{m_{f}%
}(\overrightarrow{p})\text{ }b^{\dag}_{r}(\overrightarrow{p})b_{r}
(\overrightarrow{p})+\epsilon_{m_{f}}(\overrightarrow{p})\text{ }d_{r}^{\dag
}(\overrightarrow{p})d_{r}(\overrightarrow{p}){\large ).} \label{diagham}%
\end{align}

This expression is central for the purpose \ of \ deciding about the
assignation of the creation and annihilation qualities to the introduced
operators and the type of indefinite metric of the Fock space of states.
\ \ This formula indicates that for massive modes quantities, the usual
assignation $b_{r}(\overrightarrow{p})$ and $d_{r}(\overrightarrow{p})$ \ to
annihilation and $b_{r}^{\dag}(\overrightarrow{p})$ and $d_{r}^{\dag
}(\overrightarrow{p})$ to creation \ operators, assures that the contribution
of their created particles (over a vacuum state annihilated by all the
$b_{r}(\overrightarrow{p})$ and $d_{r}(\overrightarrow{p})$) to the energy
will be bounded from below. Then, let us adopt the following commutation
relations for the massive type operators%
\begin{align}
\{b_{r}^{\dag}(\overrightarrow{p}_{1}),b_{r^{\prime}}(\overrightarrow{p}%
_{2})\}  &  =s_{m}\text{ }\delta_{r,r^{\prime}}\delta^{(K)}(\overrightarrow{p}%
_{1},\overrightarrow{p}_{2}),\\
\{d_{r}^{\dag}(\overrightarrow{p}_{1}),d_{r^{\prime}}(\overrightarrow{p}%
_{2})\}  &  =s_{m}\text{ }\delta_{r,r^{\prime}}\delta^{(K)}(\overrightarrow{p}%
_{1},\overrightarrow{p}_{2})\\
\{A,B\}  &  =AB+BA,\nonumber
\end{align}
where the number $s_{m}=\pm1$\ yet allows to select positive or negative
metric quantization of the associated states for massive fermions \cite{nagy,
nakanishi}. The function $\delta^{(K)}(\overrightarrow{p}_{1}%
,\overrightarrow{p}_{2})$\ represent the Kronecker Delta function, equal to
one for equal arguments and vanishing when they differ. In what follows, the
definition of the Dirac's Delta function of the coordinates will be employed
which is related with the discrete momenta in the quantization box as follows
\begin{align}
\delta^{(D)}(\overrightarrow{x}_{1}-\overrightarrow{x}_{2})  &  =\sum
_{\overrightarrow{p}}\frac{1}{V}\exp(i\overrightarrow{p}.(\overrightarrow{x}%
_{1}-\overrightarrow{x}_{2}))\nonumber\\
&  =\int\frac{dp^{3}}{(2\pi)^{3}}\exp(i\overrightarrow{p}.(\overrightarrow{x}%
_{1}-\overrightarrow{x}_{2})).
\end{align}

For the case of the massless modes quantities, it is clear \ that in order to
the energy be positive, given the fermion character of the fields, \ the
assignation can be to consider a creation type for the operators
$a_{l}(\overrightarrow{p})$, $c_{l}(\overrightarrow{p})$ and \ an annihilation
nature for the operators $a_{l}^{\dag}(\overrightarrow{p}),$ $c_{l}^{\dag
}(\overrightarrow{p})$\ . With this interpretation, and since the particles
are fermions we will define the adopted here creation and annihilation
operators $\widehat{a}_{l}(\overrightarrow{p}),\widehat{c}_{l}%
(\overrightarrow{p})$ and their Hermitian conjugates as
\begin{align}
\widehat{a}_{l}(\overrightarrow{p})  &  =a_{l}^{\dag}(\overrightarrow{p}%
),\text{ \ }\widehat{a}_{l}^{\dag}(\overrightarrow{p})=a_{l}%
(\overrightarrow{p}),\text{\ }\nonumber\\
\widehat{c}_{l}(\overrightarrow{p})  &  =c_{l}^{\dag}(\overrightarrow{p}%
),\text{ \ }\widehat{c}_{l}^{\dag}(\overrightarrow{p})=c_{l}%
(\overrightarrow{p}),
\end{align}
assuming to obey the anticommutation relations \
\begin{equation}
\{\widehat{a}_{l}^{\dag}(\overrightarrow{p}_{1}),\widehat{a}_{l^{\prime}%
}(\overrightarrow{p}_{2})\}=s_{0}\text{ }\delta_{l,l^{\prime}}\delta
^{(K)}(\overrightarrow{p}_{1},\overrightarrow{p}_{2}),\{\widehat{c}_{l}^{\dag
}(\overrightarrow{p}_{1}),\widehat{c}_{l^{\prime}}(\overrightarrow{p}%
_{2})\}=s_{0}\text{ }\delta_{l,l^{\prime}}\delta^{(K)}(\overrightarrow{p}%
_{1},\overrightarrow{p}_{2}),
\end{equation}
where in addition, any anticommutator between any two elements pertaining to
the set formed by the operators and their Hermitian transposes, vanishes, if
the two operators pertain to different classes of the four types
$\ "a\,\,","c\,\,","b\,\,"$ \ and "$d\,\,".$ \ The number $s_{0}=\pm1$ in this
case, also allows for fixing positive or negative norm for the massless single
particle states. Therefore, in what follows, the quantities $\widehat{a}%
_{l}(\overrightarrow{p})$, $\widehat{c}_{l}(\overrightarrow{p}),$
$b_{r}(\overrightarrow{p})$ and $d_{r}(\overrightarrow{p})$ will be assumed to
define the annihilation operators of the massless and massive quarks, and
their Hermitian conjugate quantities are associated to the creation operators
of the corresponding quarks.

Considering these definitions \ $H$ \ can be expressed in the traditional
form
\begin{align}
H  &  =\sum_{\overrightarrow{p}}\sum_{l=\pm1}{\large (}\epsilon_{0}%
(\overrightarrow{p})\text{ }\widehat{a}_{_{l}}^{\dag}(\overrightarrow{p}%
)\widehat{a}_{l}(\overrightarrow{p})+\epsilon_{0}(\overrightarrow{p})\text{
}\widehat{c}_{l}^{\dag}(\overrightarrow{p})\widehat{c}_{l}(\overrightarrow{p}%
){\large )}+\nonumber\\
&  \sum_{\overrightarrow{p}}\sum_{r=1,2}{\large (}\epsilon_{m_{f}%
}(\overrightarrow{p})\text{ }b_{r}^{\dag}(\overrightarrow{p})b_{r}%
(\overrightarrow{p})+\epsilon_{m_{f}}(\overrightarrow{p})\text{ }d_{r}^{\dag
}(\overrightarrow{p})d_{r}(\overrightarrow{p}){\large ).} \label{hamfin}%
\end{align}

The fields are now written as follows \
\begin{align}
\Psi(x)  &  =\Psi_{m}(x)+\Psi_{0}(x)=\sum_{\kappa=0,m}\Psi_{\kappa}(x),\\
\Psi_{0}(x)  &  =\frac{1}{\sqrt{V}}\sum_{\overrightarrow{p}}\sum_{l=\pm
1}{\Large (}\widehat{a}_{l}^{\dag}(\overrightarrow{p})u^{l}_{0}%
(\overrightarrow{p})\exp(-i\text{ }(\epsilon_{0}(\overrightarrow{p}%
)t-\overrightarrow{p}.\overrightarrow{x}))+\nonumber\\
&  \text{ \ \ \ \ \ \ \ \ \ \ \ \ \ \ \ \ \ \ \ \ \ \ \ \ \ \ \ }%
\widehat{c}_{l}(\overrightarrow{p})v^{l}_{0}(\overrightarrow{p})\exp(i\text{
}(\epsilon_{0}(\overrightarrow{p})t+\overrightarrow{p}.\overrightarrow{x}%
)){\Large )}\nonumber\\
&  =u_{0}(x)+v_{0}(x),\\
\Psi_{m}(x)  &  =\frac{1}{\sqrt{V}}\sum_{\overrightarrow{p}}\sum
_{r=1,2}{\Large (}b_{r}(\overrightarrow{p})u_{m_{f}}^{r}(\overrightarrow{p}%
)\exp(-i\text{ }(\epsilon_{m_{f}}(\overrightarrow{p})t-\overrightarrow{p}%
.\overrightarrow{x}))+\nonumber\\
&  \text{ \ \ \ \ \ \ \ \ \ \ \ \ \ \ \ \ \ \ \ \ \ \ \ \ \ \ \ }d_{r}^{\dag
}(\overrightarrow{p})v_{m_{f}}^{r}(\overrightarrow{p})\exp(i\text{ }%
(\epsilon_{m_{f}}(\overrightarrow{p})t+\overrightarrow{p}.\overrightarrow{x}%
)){\Large )}\\
&  =u_{m}(x)+v_{m}(x).\nonumber
\end{align}

Now, in order to further define the quantization, it can be also taken into
account that in the considered free theory, the expression $\ $for
$H\ $\ should also be consistent \ with the assumed classical time dependence
of the modes, which was employed to write the time derivatives of the \ field
defining $H$. \ For the massless case, this condition imposes that the time
evolution of the operators generated by the usual similarity transformation
with the evolution operator of the operators $\widehat{a}_{l}%
(\overrightarrow{p})$ and $\widehat{c}_{l}^{\dag}(\overrightarrow{p})$ should
reproduce the classical time dependence of the field $\Psi_{0}(x)$. That is
\begin{align}
\widehat{a}_{l}(\overrightarrow{p},t)  &  \equiv\exp(i\text{ }H\text{
}t)\text{ }\widehat{a}_{l}(\overrightarrow{p})\exp(-iH\text{ }t).\nonumber\\
&  =\exp(i\text{ }\epsilon_{0}(\overrightarrow{p})\text{ }\widehat{a}%
_{l}^{\dag}(\overrightarrow{p})\widehat{a}_{l}(\overrightarrow{p})\text{
}t)\text{ }\widehat{a}_{l}(\overrightarrow{p})\exp(-i\text{ }\epsilon
_{0}(\overrightarrow{p})\text{ }\widehat{a}_{l}^{\dag}(\overrightarrow{p}%
)\widehat{a}_{l}(\overrightarrow{p})\text{ }t).\nonumber\\
&  =\exp(-s_{0}\text{ }i\text{ }\epsilon_{0}(\overrightarrow{p})\text{
}t)\text{ }\widehat{a}_{l}(\overrightarrow{p})\rightarrow\exp(i\text{
}\epsilon_{0}(\overrightarrow{p})\text{ }t)\text{ }\widehat{a}_{l}%
(\overrightarrow{p})\\
\widehat{c}_{l}^{\dag}(\overrightarrow{p},t)=  &  \exp(i\text{ }H\text{
}t)\text{ }\widehat{c}_{l}^{\dag}(\overrightarrow{p})\text{ }\exp(-i\text{
}H\text{ }t).\nonumber\\
&  \exp(i\text{ }\epsilon_{0}(\overrightarrow{p})\text{ }\widehat{c}_{l}%
^{\dag}(\overrightarrow{p})\widehat{c}_{l}(\overrightarrow{p})\text{ }t)\text{
}\widehat{c}_{l}^{\dag}(\overrightarrow{p})\text{ }\exp(-i\text{ }\epsilon
_{0}(\overrightarrow{p})\text{ }\widehat{c}_{l}^{\dag}(\overrightarrow{p}%
)\widehat{c}_{l}(\overrightarrow{p})t).\nonumber\\
&  =\exp(i\text{ }s_{0}\epsilon_{0}(\overrightarrow{p})\text{ }t)\text{
}\widehat{c}_{l}^{\dag}(\overrightarrow{p})\rightarrow\exp(-i\text{ }%
\epsilon_{0}(\overrightarrow{p})\text{ }t)\text{ }\widehat{c}_{l}%
(\overrightarrow{p},0).
\end{align}

Thus, the Heisenberg time evolution of the operators will coincide with the
classical one, upon the selection $s_{0}=-1.$ This indicates that the massless
field states should have negative norms.

On the contrary, the similar consistency conditions directly lead to the usual
selection $s_{m}=1$ determining the usual positive norm for the one particle
massive states. \ \ We thus conclude that, as suggested by the \ initial form
of the free propagator, the Fock space of the free theory \ show an indefinite
metric for fermion \ states, \ in which massless one particle states have
negative norm and massive states \ show a positive one. \

\subsection{ The operator field equations}

Let us show that the defined field operator and its Hermitian conjugate solve
the quantum version of the classical Hamiltonian and Lagrange equations
following by applying the standard quantization rules to the generalized
Hamiltonian system.

Since the entering two fields in any of the following expressions, are defined
by two sets of creation or annihilation operators with a \ vanishing
anticommutator between any element in one of the sets with any element
pertaining to the other set, \ the following standard vanishing commutator
relation at different spatial point and coincident time, follow for the
massless and massive components of the fields \
\begin{align}
\{\Psi_{0}(x),\Psi_{0}(x^{\prime})\}|_{x_{0}=x_{0}^{\prime}}  &
=0,\nonumber\\
\{\Psi_{0}(x),\Psi_{m}(x^{\prime})\}|_{x_{0}=x_{0}^{\prime}}  &
=0,\nonumber\\
\{\Psi_{0}(x),\Psi_{m}^{\dag}(x^{\prime})\}|_{x_{0}=x_{0}^{\prime}}  &
=0,\nonumber\\
\{\Psi_{m}(x),\Psi_{m}(x^{\prime})\}|_{x_{0}=x_{0}^{\prime}}  &  =0.
\end{align}

Their Hermitian conjugate expressions \ furnish a similar set of relations
including Hermitian conjugate fields. \ \ \ For the relation between the
fields and their \ Hermitian conjugate (h.c.) being of the same massless or
massive class, it follows%
\begin{align}
\{\Psi_{0}(x),\Psi_{0}^{\dag}(x^{\prime})\}|_{x_{0}=x_{0}^{\prime}} &
=T_{1}+T_{2},\nonumber\\
T_{1} &  =-\frac{1}{V}\sum_{\overrightarrow{p},l}\sum_{\overrightarrow{p}%
^{\prime},l^{\prime}}u_{l}(p)\otimes u_{l^{\prime}}^{\dag}(\overrightarrow{p}%
^{\prime})\text{ }\{\widehat{a}_{l^{\prime}}^{\dag}(\overrightarrow{p}%
^{\prime}),\widehat{a}_{l}(\overrightarrow{p})\}\times\nonumber\\
&  \exp{\large (}i\text{ }(\epsilon_{0}(\overrightarrow{p}^{\prime}%
)-\epsilon_{0}(\overrightarrow{p}))x_{0}-i(\overrightarrow{p}^{\prime
}.\overrightarrow{x}-\overrightarrow{p}.\overrightarrow{x}){\large )}%
\nonumber\\
&  =-\frac{1}{V}\sum_{\overrightarrow{p},l}u_{l}(p)u_{l}^{\dag}%
(\overrightarrow{p})\exp{\large (}-i\overrightarrow{p}.(\overrightarrow{x}%
^{\prime}-\overrightarrow{x}){\large )}\nonumber\\
T_{2} &  =-\frac{1}{V}\sum_{\overrightarrow{p},l}v_{l}(p)\otimes v_{l}^{\dag
}(\overrightarrow{p})\exp{\large (}-i\overrightarrow{p}.(\overrightarrow{x}%
^{\prime}-\overrightarrow{x}){\large ),}\nonumber\\
\{\Psi_{0}(x),\Psi_{0}^{\dag}(x^{\prime})\}|_{x_{0}=x_{0}^{\prime}} &
=-\text{ }\delta^{(D)}(\overrightarrow{x}-\overrightarrow{x}^{\prime})I,
\end{align}
in which the term $T_{2}$ is following after similar steps. The symbol
$\otimes$ will indicate the external product of two spinors leading to a
spinor matrix. The negative metric anticommutation relation and spinor
completeness relations had been employed%
\begin{align}
\{\widehat{a}_{l^{\prime}}^{\dag}(\overrightarrow{p}^{\prime}),\widehat{a}%
_{l}(\overrightarrow{p})\} &  =-\text{ }\delta_{l,l^{\prime}}\text{ }%
\delta^{(K)}(\overrightarrow{p}^{\prime},\overrightarrow{p})\text{ }I,\\
\{\widehat{c}_{l^{\prime}}^{\dag}(\overrightarrow{p}^{\prime}),\widehat{c}%
_{l}(\overrightarrow{p})\} &  =-\text{ }\delta_{l,l^{\prime}}\text{ }%
\delta^{(K)}(\overrightarrow{p}^{\prime},\overrightarrow{p})\text{ }I,\\
I &  =\sum_{l=\pm1}(\text{ }u_{l}(p)\otimes u_{l}^{\dag}(\overrightarrow{p}%
)+v_{l}(p)\otimes v_{l}^{\dag}(\overrightarrow{p})\text{ }),\\
\delta^{(D)}(\overrightarrow{x}-\overrightarrow{x}^{\prime}) &  =\sum
_{\overrightarrow{p}}\frac{1}{V}\exp{\large (}-i\text{ }\overrightarrow{p}%
.(\overrightarrow{x}^{\prime}-\overrightarrow{x}){\large )}\nonumber\\
&  \simeq\int\frac{dp^{3}}{(2\pi)^{3}}\exp{\large (}-i\text{ }%
\overrightarrow{p}.(\overrightarrow{x}^{\prime}-\overrightarrow{x}%
){\large ).}\nonumber
\end{align}

The completeness condition follows from the massless spinor definitions in
(\ref{massless1}) and (\ref{massless2}). Similarly, for \ the commutation
relations between the massive field and its h.c. it is possible to write
\begin{align}
\{\Psi_{m}(x),\Psi_{m}^{\dag}(x^{\prime})\}|_{x_{0}=x_{0}^{\prime}} &
=S_{1}+S_{2},\nonumber\\
S_{1} &  =\frac{1}{V}\sum_{\overrightarrow{p},\text{ }r=1,2}\sum
_{\overrightarrow{p}^{\prime},r^{\prime}=1,2}u_{r}(p)\otimes u_{r^{\prime}%
}^{\dag}(\overrightarrow{p}^{\prime})\text{ }\{b_{r^{\prime}}^{\dag
}(\overrightarrow{p}^{\prime}),b_{r}(\overrightarrow{p})\}\times\nonumber\\
&  \exp{\large (}i\text{ }(\epsilon_{m_{f}}(\overrightarrow{p}^{\prime
})-\epsilon_{m_{f}}(\overrightarrow{p}))x_{0}-i\text{ }(\overrightarrow{p}%
^{\prime}.\overrightarrow{x}^{\prime}-\overrightarrow{p}.\overrightarrow{x}%
){\large )}\nonumber\\
&  =\frac{1}{V}\sum_{\overrightarrow{p},\text{ }r=1,2}u_{r}(p)\otimes
u_{r}^{\dag}(\overrightarrow{p})\exp{\large (}-i\text{ }\overrightarrow{p}%
.(\overrightarrow{x}^{\prime}-\overrightarrow{x}){\large ),}\nonumber\\
S_{2} &  =\frac{1}{V}\sum_{\overrightarrow{p},r=1,2}v_{r}(p)\otimes
v_{r}^{\dag}(\overrightarrow{p})\exp{\large (}-i\text{ }\overrightarrow{p}%
.(\overrightarrow{x}^{\prime}-\overrightarrow{x}){\large ),}\nonumber\\
\{\Psi_{m}(x),\Psi_{m}^{\dag}(x^{\prime})\}|_{x_{0}=x_{0}^{\prime}} &
=\delta^{(D)}(\overrightarrow{x}-\overrightarrow{x}^{\prime})\text{ }I.
\end{align}

In this massive case, the \ relations employed to arrive to the expression of
the commutator were%
\begin{align}
\{b_{r^{\prime}}^{\dag}(\overrightarrow{p}^{\prime}),b_{r}(\overrightarrow{p}%
)\} &  =\delta_{r,r^{\prime}}\delta^{(K)}(\overrightarrow{p}^{\prime
},\overrightarrow{p})\text{ }I,\\
\{d_{r^{\prime}}^{\dag}(\overrightarrow{p}^{\prime}),d_{r}(\overrightarrow{p}%
)\} &  =\delta_{r,r^{\prime}}\delta^{(K)}(\overrightarrow{p}^{\prime
},\overrightarrow{p})\text{ }I,\\
I &  =\sum_{r=\pm1}(u_{r}(p)\otimes u_{r}^{\dag}(\overrightarrow{p}%
)+v_{r}(p)\otimes v_{r}^{\dag}(\overrightarrow{p})).
\end{align}

The obtained commutation relations between the massless and massive components
of the fields, then allow to write for the total fields $\Psi$ and $\Psi
^{\dag}$, the commutations relations
\begin{align}
\{\Psi(x),\Psi(x^{\prime})\}|_{x_{0}=x_{0}^{\prime}}  &  =\{\Psi^{\dag
}(x),\Psi^{\dag}(x^{\prime})\}|_{x_{0}=x_{0}^{\prime}}=0,\\
\{\Psi(x),\Psi^{\dag}(x^{\prime})\}|_{x_{0}=x_{0}^{\prime}}  &  =\{\Psi
_{0}(x),\Psi_{0}^{\dag}(x^{\prime})\}|_{x_{0}=x_{0}^{\prime}}+\{\Psi
_{m}(x),\Psi_{m}^{\dag}(x^{\prime})\}|_{x_{0}=x_{0}^{\prime}}\\
&  =0,\nonumber
\end{align}
which reproduce the similar \ relations between the Poisson brackets between
the coordinates in the classical theory.

Let us now represent the momenta operators defined in terms of the operators
$\Psi(x),\Psi^{\dag}(x)$ by the same classical expressions \
\begin{align}
\Pi_{\Psi}(x) &  =-\frac{i}{2}\Psi^{\dagger}(x)+\varkappa\text{ }\partial
_{t}\Psi^{\dagger}(x)\beta,\\
\Pi_{\Psi^{\dagger}}(x) &  =-\frac{i}{2}\Psi(x)-\varkappa\text{ }\beta\text{
}\partial_{t}\Psi(x),
\end{align}
written in terms of the defined field operators. Then, for the anticommutator
of them with the "coordinates" $\Psi(x)$ and $\Psi(x)$ \ it follows
\begin{align}
\{\Psi(x),\Pi_{\Psi}(x^{\prime})\}|_{x_{0}=x_{0}^{\prime}} &  =\varkappa\text{
}\{\Psi_{0}(x),\text{ }\partial_{t}\Psi_{0}^{\dagger}(x^{\prime}%
)\beta\}|_{x_{0}=x_{0}^{\prime}}+\varkappa\text{ }\{\Psi_{m}(x),\partial
_{t}\Psi_{m}^{\dag}(x^{\prime})\beta\}|_{x_{0}=x_{0}^{\prime}}\nonumber\\
&  =R_{1}+R_{2},\nonumber\\
R_{1} &  =-\frac{i}{V}\text{ }\varkappa\sum_{\overrightarrow{p},l}\epsilon
_{0}(\overrightarrow{p}){\large (}u_{l}(p)\otimes u_{l}^{\dag}%
(\overrightarrow{p})-v_{l}(p)\otimes v_{l}^{\dag}(\overrightarrow{p}%
){\large )}\exp{\large (}-i\text{ }\overrightarrow{p}.(\overrightarrow{x}%
^{\prime}-\overrightarrow{x}){\large ),}\nonumber\\
R_{2} &  =\frac{i}{V}\text{ }\varkappa\sum_{\overrightarrow{p},r}\epsilon
_{m}(\overrightarrow{p}){\large (}u_{r}(p)\otimes u_{r}^{\dag}%
(\overrightarrow{p})-v_{r}(p)\otimes v_{r}^{\dag}(\overrightarrow{p}%
){\large )}\exp{\large (}-i\text{ }\overrightarrow{p}.(\overrightarrow{x}%
^{\prime}-\overrightarrow{x}){\large ).}%
\end{align}

But, after employing the spinors definitions (\ref{massive1}), (\ref{massive2}%
) and (\ref{massless1}), (\ref{massless2}) it follows
\begin{align}
\sum_{l=\pm1}(u_{l}(p)\otimes u_{l}^{\dag}(\overrightarrow{p})-v_{l}(p)\otimes
v_{l}^{\dag}(\overrightarrow{p})) &  \equiv\left(
\begin{array}
[c]{cc}%
0 & \frac{\overrightarrow{\sigma}.\overrightarrow{p}}{|\overrightarrow{p}|}\\
\frac{\overrightarrow{\sigma}.\overrightarrow{p}}{|\overrightarrow{p}|} & 0
\end{array}
\right)  ,\\
\sum_{r=1,2}(u_{r}(p)\otimes u_{r}^{\dag}(\overrightarrow{p})-v_{r}(p)\otimes
v_{r}^{\dag}(\overrightarrow{p})) &  \equiv\frac{m_{f}}{\epsilon_{m_{f}%
}(\overrightarrow{p})}\left(
\begin{array}
[c]{cc}%
I & \frac{\overrightarrow{\sigma}.\overrightarrow{p}}{m_{f}}\\
\frac{\overrightarrow{\sigma}.\overrightarrow{p}}{m_{f}} & -I
\end{array}
\right)  ,
\end{align}
which gives the commutation rule
\begin{equation}
\{\Psi(x),\Pi_{\Psi}(x^{\prime})\}|_{x_{0}=x_{0}^{\prime}}=i\text{ }%
\delta^{(D)}(\overrightarrow{x}-\overrightarrow{x}^{\prime})I.\label{commu1}%
\end{equation}

Therefore, the momentum operators defined in terms of the field operators as
following from the classical equations also satisfies the usual quantization
condition \ defining the anticommutator between \ the momenta and coordinates
as given by the classical Poisson bracket between them times the imaginary unit.

Considering that in identical form as it happens for the classical Grassman
fields and momenta $(\Pi_{\Psi}(x^{\prime}))^{\dag}=-\Pi_{\Psi^{\dag}%
}(x^{\prime}),$ and after taking the Hermitian transpose of \ (\ref{commu1})
\ it also follows%
\begin{equation}
\{\Psi^{\dag}(x),\Pi_{\Psi^{\dag}}(x^{\prime})\}|_{x_{0}=x_{0}^{\prime}%
}=i\text{ }\delta^{(D)}(\overrightarrow{x}-\overrightarrow{x}^{\prime})I,
\end{equation}
for the other set of conjugate pair of coordinate and momentum in the system.

\ The anticommutation between momenta operators associated to the same fields
$\Psi$ or $\Psi^{\dag}$\
\begin{equation}
\{\Pi_{\Psi}(x),\Pi_{\Psi}(x^{\prime})\}|_{x_{0}=x_{0}^{\prime}}=\{\Pi
_{\Psi^{\dag}}(x),\Pi_{\Psi^{\dag}}(x^{\prime})=0,
\end{equation}
directly follow because any of the creation or annihilation operators included
in the set $S$ of them defining each of the entering momenta does not has its
Hermitian transposed included in this same set $S$.

Further, after employing the definitions of the momenta in terms of the
fields, the anticommutator between momenta operators associated to the fields
$\Psi$ and $\Psi^{\dag}$ can be expressed \ in the form \
\begin{align}
\{\Pi_{\Psi}(x),\Pi_{\Psi^{\dag}}(x^{\prime})\}|_{x_{0}=x_{0}^{\prime}}  &
=-\varkappa^{2}\{\partial_{t}\Psi^{\dag}(x^{\prime})\text{ }\beta,\beta\text{
}\partial_{t}\Psi(x)\}+\nonumber\\
&  \frac{i\text{ }\varkappa}{2}(\{\Psi^{\dag}(x^{\prime}),\beta\text{
}\partial_{t}\Psi(x)\}-\{\partial_{t}\Psi^{\dag}(x^{\prime})\text{ }\beta
,\Psi(x)\})\nonumber\\
&  =-\varkappa^{2}\{\partial_{t}\Psi_{0}^{\dag}(x^{\prime})\text{ }\beta
,\beta\text{ }\partial_{t}\Psi_{0}(x)\}-\varkappa^{2}\{\partial_{t}\Psi
_{m}^{\dag}(x^{\prime})\text{ }\beta,\beta\text{ }\partial_{t}\Psi
_{m}(x)\}+\nonumber\\
&  -\frac{i\text{ }}{2}\{\Psi^{\dag}(x^{\prime}),\Pi_{\Psi^{\dag}}%
(x)\}-\frac{i\text{ }}{2} \{\Pi_{\Psi}(x^{\prime}),\Psi(x)\}.
\end{align}
But, the direct evaluation \ leads to
\begin{align}
-\varkappa^{2}\{\partial_{t}\Psi_{0}^{\dag}(x^{\prime})\beta,\beta\partial
_{t}\Psi_{0}(x)\}  &  =\frac{I}{V}\sum_{\overrightarrow{p}}\frac
{|\overrightarrow{p}|^{2}}{m_{f}^{2}}\exp(i\text{ }\overrightarrow{p}%
.(\overrightarrow{x}-\overrightarrow{x}^{\prime})),\\
-\varkappa^{2}\{\partial_{t}\Psi_{m}^{\dag}(x^{\prime})\beta,\beta\partial
_{t}\Psi_{m}(x)\}  &  =-\frac{I}{V}\sum_{\overrightarrow{p}}\frac
{(|\overrightarrow{p}|^{2}+m_{f}^{2})}{m_{f}^{2}}\exp(i\text{ }%
\overrightarrow{p}.(\overrightarrow{x}-\overrightarrow{x}^{\prime})),\\
-\frac{i\text{ }}{2}\{\Psi^{\dag}(x^{\prime}),\Pi_{\Psi^{\dag}}(x)\}-\frac
{i\text{ }}{2}\{\Pi_{\Psi}(x^{\prime}),\Psi(x) \}  &  =\frac{I}{V}%
\sum_{\overrightarrow{p}}\exp(i\text{ }\overrightarrow{p}.(\overrightarrow{x}%
-\overrightarrow{x}^{\prime})),
\end{align}
which after substituted show the resting standard anticommutation rules%
\begin{equation}
\{\Pi_{\Psi}(x),\Pi_{\Psi^{\dag}}(x^{\prime})\}|_{x_{0}=x_{0}^{\prime}}=0.
\end{equation}

Therefore, the new defined quantum fields $\Psi(x)$\ and $\Psi^{\dag}(x)$ and
their momenta $\Pi_{\Psi}(x)$ and $\Pi_{\Psi^{\dagger}}$ satisfy \ the
standard quantization rules of the classical classical Grassman counterparts.
The Hamiltonian of the system $H$ can be now defined in terms of the classical
one by substituting the classical fields and canonical momenta by the above
defined operators in a way assuring that $H$ is hermitian. The same classical
expression gives a Hermitian representation in the form
\begin{align}
H  &  =\int dx^{3}(\Pi_{\Psi}(x)\partial_{t}\Psi(x)+\Pi_{\Psi^{\dagger}%
}(x)\partial_{t}\Psi^{\dagger}(x))-L\nonumber\\
&  =\int dx^{3}{\huge \{}\frac{1}{x}{\Large (}-\Pi_{\Psi}(x)\text{ }%
\beta\text{\ }\Pi_{\Psi^{\dagger}}(x)+\frac{1}{4}\Psi^{\dagger}(x)\text{
}\beta\text{ }\Psi(x)-\nonumber\\
&  \frac{i}{2}\Psi^{\dagger}(x)\text{ }\beta\text{ }\Pi_{\Psi^{\dagger}%
}(x)-\frac{i}{2}\Pi_{\Psi}(x)\text{ }\beta\text{ }\Psi(x){\Large )}%
+\nonumber\\
&  \text{\ }\frac{i}{2}\Psi^{\dagger}(x)\alpha^{i}\overleftrightarrow{\partial
}_{i}\Psi(x)+\varkappa\text{ }\Psi^{\dagger}(x)\overleftarrow{\nabla
}.\overrightarrow{\nabla}\beta\Psi(x){\huge \}}.
\end{align}

Then, the use of the commutation relations
\begin{align}
\{\Psi^{\dag}(x),\Pi_{\Psi^{\dag}}(x^{\prime})\}|_{x_{0}=x_{0}^{\prime}}  &
=i\text{ }\delta^{(D)}(\overrightarrow{x}-\overrightarrow{x}^{\prime})\text{
}I,\\
\{\Psi(x),\Pi_{\Psi}(x^{\prime})\}|_{x_{0}=x_{0}^{\prime}}  &  =i\text{
}\delta^{(D)}(\overrightarrow{x}-\overrightarrow{x}^{\prime})\text{ }I,
\end{align}
in conjunction with the rest of them among the fields and momenta which
vanish, and the \ Heisenberg equation of motions for arbitrary operators $A$%
\begin{equation}
\partial_{t}A(t)=i\text{ }\left[  H,A(t)\right]  ,
\end{equation}
permits to check that the fields and momenta obey the \ equations of motion%
\begin{align}
\partial_{t}\Psi(x)  &  =-\frac{1}{\varkappa}\beta\text{ }(\Pi_{\Psi^{\dagger
}}(x)+\frac{i}{2}\Psi(x)),\\
\partial_{t}\Psi^{\dagger}  &  =\frac{1}{\varkappa}(\Pi_{\Psi}(x)+\frac{i}%
{2}\Psi^{\dagger}(x))\text{ }\beta,\\
\partial_{t}\Pi_{\Psi}(x)\text{ }  &  =\frac{i}{2\varkappa}\Pi_{\Psi}(x)\text{
}\beta-\frac{1}{4\varkappa}\Psi^{\dagger}(x)\text{ }\beta\\
&  +i\text{ }\Psi^{\dagger}\alpha^{i}\overleftarrow{\partial}_{i}%
+\varkappa\nabla^{2}\Psi^{\dagger}\beta,\\
\partial_{t}\Pi_{\Psi^{\dagger}}(x)  &  =-\frac{i}{2\varkappa}\beta\text{ }%
\Pi_{\Psi^{\dagger}}(x)+\frac{1}{4\varkappa}\beta\text{ }\Psi(x)\nonumber\\
&  +i\text{ }\alpha^{i}\partial_{i}\Psi^{\dagger}-\varkappa\text{ }\beta\text{
}\nabla^{2}\Psi.
\end{align}
The further elimination of the momenta among these equations check that the
"coordinate" fields also satisfy the original Lagrange equations%
\begin{align}
(-i\text{ }\gamma^{\mu}\partial_{\mu}-\varkappa\text{ }\partial^{2})\text{
}\Psi(x)  &  =0,\\
\Psi^{\dag}(x)\text{ }(i\text{ }\gamma^{\mu}\overleftarrow{\partial}_{\mu
}-\varkappa\text{ }\overleftarrow{\partial}^{2})  &  =0.
\end{align}

\section{Summmary}

In this work, it has been studied the nature of the free quantization of the
quark free Lagrangian associated to the recently proposed local and gauge
invariant alternative for QCD for interacting massive quarks. It was argued
that the free theory can be quantized in a way in which the massive quarks
states can show a positive metric and the massless ones a negative norm. The
quantization of the classical theory had been implemented, and an explicit
representation of the fields and momenta operators solving the Heisenberg
equations of motions was constructed. It is underlined that the accepted
absence of asymptotic states for quark and gluons in QCD indicates that the
appearance of indefinite metric in the quark sector of the theory does not
necessarily present a definite limitation of the the investigated alternative
for QCD to describe nature. This conclusion follows because, precisely, the
physical validity of the Lee-Wick theories rests in the absence of stable
asymptotic states associated with the negative metric states due to the
interactions \cite{lee-wick}. Therefore, the recognized today absence of free
quarks and gluons in QCD can imply the satisfaction of this condition, through
the confinement effect. The discussion also allowed to define a representation
for the general classical action, and of the associated Dirac propagator of
the Feynman expansion, which could improve the previous ones employed in Ref.
\cite{mQCD}, for applications to high energy processes, where the massive
quark states show approximate existence due to asymptotic freedom. It also
could be expected to be of help in Nuclear QCD studies were also massive quark
are relevant.

\begin{acknowledgments}
The author would like to acknowledge \ Justo Lopez, Maximo Porrati, \ Benjamin
Grinstein and Manuel Asorey, for communicating their views about the possible
connections of the modified QCD being investigated with Lee-Wick \ theories. I
am also indebted by the helpful support received from various institutions:
the Caribbean Network on Quantum Mechanics, Particles and Fields (Net-35) of
the ICTP Office of External Activities (OEA), the "Proyecto Nacional de
Ciencias B\'{a}sicas" (PNCB) of CITMA, Cuba.
\end{acknowledgments}

\appendix

\section{}

\ In this appendix we sketch the \ derivation of the terms which define the
standard formula for the \ energy operator as the superposition of the number
operators for the various particles after multiplied by their respective
energies. Below, the main algebraic steps \ in evaluating the general
components of the energy operator \ (\ref{ham2}) are explicitly illustrated
for two components. The evaluation for rest of the terms follows very similar
algebraical lines.

For the contribution of massless positive energy states it follows\ \ \
\begin{align*}
\int dx^{3}u_{0}^{\dagger}(x)h(\partial)u_{0}(x)  &  =\sum_{\overrightarrow{p}%
,}\sum_{\overrightarrow{p}^{\prime}}\sum_{l=\pm1}\sum_{l^{\prime}=\pm1}%
\frac{1}{2}\times\\
&  \left(
\begin{array}
[c]{cc}%
\beta^{\dag l^{\prime}}(\overrightarrow{p}^{\prime}) & l^{\prime}\beta^{\dag
l^{\prime}}(\overrightarrow{p}^{\prime})
\end{array}
\right)  \times\\
&  \left(
\begin{array}
[c]{cc}%
\varkappa(\epsilon_{0}(\overrightarrow{p}^{\prime})\text{ }\epsilon
_{0}(\overrightarrow{p})+\overrightarrow{p}^{\prime}.\overrightarrow{p}) &
\overrightarrow{\sigma}.\overrightarrow{p}\\
\overrightarrow{\sigma}.\overrightarrow{p} & -\varkappa(\epsilon
_{0}(\overrightarrow{p}^{\prime})\epsilon_{0}(\overrightarrow{p}%
)+\overrightarrow{p}^{\prime}.\overrightarrow{p})
\end{array}
\right)  \left(
\begin{array}
[c]{c}%
\beta^{l}(\overrightarrow{p})\\
l\text{ }\beta^{l}(\overrightarrow{p})
\end{array}
\right)  \times\\
&  \frac{1}{V}\int dx^{3}\exp(i\text{ }(\overrightarrow{p}-\overrightarrow{p}%
^{\prime}).\overrightarrow{x})\text{ }\widehat{a}_{l^{\prime}}^{\dag
}(\overrightarrow{p}^{\prime})\widehat{a}_{l}(\overrightarrow{p})
\end{align*}

\begin{align}
\text{ \ \ \ \ \ \ \ \ \ \ \ \ \ \ \ \ \ \ \ }  &  =\sum_{\overrightarrow{p}%
,}\sum_{\overrightarrow{p}^{\prime}}\sum_{l=\pm1}\sum_{l^{\prime}=\pm1}%
\delta^{l\text{ }l^{\prime}}\delta^{(K)}(\overrightarrow{p},\overrightarrow{p}%
^{\prime})\frac{1}{2}(\beta^{l}(\overrightarrow{p}))^{\dagger}\beta
^{l}(\overrightarrow{p}){\large (}\varkappa(\epsilon_{0}(\overrightarrow{p}%
)\epsilon_{0}(\overrightarrow{p})+\nonumber\\
&  \text{ \ \ \ \ \ }\overrightarrow{p}.\overrightarrow{p})(1-l\text{
}l^{\prime})+l(l+l^{\prime})|\overrightarrow{p}|{\Large )}\widehat{a}%
_{l^{\prime}}^{\dag}(\overrightarrow{p}^{\prime})\widehat{a}_{l}%
(\overrightarrow{p})\nonumber\\
&  =\sum_{\overrightarrow{p}}\sum_{l=\pm1}\frac{1}{2}\text{ }2\text{ }%
l^{2}|\overrightarrow{p}|\widehat{a}_{l}^{\dag}(\overrightarrow{p}%
)\widehat{a}_{l}(\overrightarrow{p})\nonumber\\
&  =\sum_{\overrightarrow{p}}\sum_{l=\pm1}\text{ }\epsilon_{0}%
(\overrightarrow{p})\widehat{a}_{l}^{\dag}(\overrightarrow{p})\widehat{a}%
_{l}(\overrightarrow{p}).
\end{align}
In a very similar way for the negative energy contributions
\begin{equation}
\int dx^{3}v_{0}^{\dagger}(x)h(\partial)v_{0}(x)=\sum_{\overrightarrow{p}}%
\sum_{l=\pm1}\text{ }\epsilon_{0}(\overrightarrow{p})\text{ }c_{l}^{\dag
}(\overrightarrow{p})c_{l}(\overrightarrow{p})
\end{equation}
In the case of the terms related with massive quarks, the derivation of the
energy term \ is sketched as follows%
\begin{align*}
\int dx^{3}u_{m}^{\dagger}(x)\text{ }h(\partial)\text{ }u_{m}(x)  &
=\sum_{\overrightarrow{p},}\sum_{\overrightarrow{p}^{\prime}}\sum_{r=1,2}%
\sum_{r^{\prime}=1,2}\sqrt{\frac{\epsilon_{m_{f}}(\overrightarrow{p}^{\prime
})+m_{f}}{2\epsilon_{m_{f}}(\overrightarrow{p}^{\prime})}}\times\sqrt
{\frac{\epsilon_{m_{f}}(\overrightarrow{p})+m_{f}}{2\epsilon_{m_{f}%
}(\overrightarrow{p})}}\\
&  \frac{1}{V}\int dx^{3}\exp(i\text{ }(\overrightarrow{p}-\overrightarrow{p}%
^{\prime}).\overrightarrow{x})\text{ }\times b_{r^{\prime}}^{\dag
}(\overrightarrow{p}^{\prime})\text{ }b_{r}(\overrightarrow{p})\times\\
&  \left(
\begin{array}
[c]{cc}%
u^{\dag r^{\prime}}(\overrightarrow{p}^{\prime})\text{ \ \ }, & u^{\dag
r^{\prime}}(\overrightarrow{p}^{\prime})\text{ }\frac{\overrightarrow{\sigma
}.\overrightarrow{p}^{\prime}}{\epsilon_{m_{f}}(\overrightarrow{p}^{\prime
})+m_{f}}%
\end{array}
\right)  \times\\
&  \left(
\begin{array}
[c]{cc}%
\varkappa\text{ }(\epsilon_{m_{f}}(\overrightarrow{p}^{\prime})\epsilon
_{m_{f}}(\overrightarrow{p})+\overrightarrow{p}^{\prime}.\overrightarrow{p}) &
-\overrightarrow{\sigma}.\overrightarrow{p}\\
-\overrightarrow{\sigma}.\overrightarrow{p} & -\varkappa\text{ }(\epsilon
_{0}(\overrightarrow{p}^{\prime})\epsilon_{0}(\overrightarrow{p}%
)+\overrightarrow{p}^{\prime}.\overrightarrow{p})
\end{array}
\right)  \times\\
&  \left(
\begin{array}
[c]{c}%
u^{r}(\overrightarrow{p})\\
\frac{\overrightarrow{\sigma}.\overrightarrow{p}}{\epsilon_{m_{f}%
}(\overrightarrow{p})+m_{f}}u^{r}(\overrightarrow{p})u^{r}(\overrightarrow{p})
\end{array}
\right)  .
\end{align*}

After evaluating the spinor products, expressing the spatial integral as the
Kronecker delta of the two spatial momenta, and using the orthonormality of
the two component spinors, this expression can be simplified in the way
described below%
\begin{align}
\int dx^{3}u_{m}^{\dagger}(x)\text{ }h(\partial)\text{ }u_{m}(x)  &
=\sum_{\overrightarrow{p},}\sum_{r=1,2}{\Large (}\varkappa(\epsilon_{m_{f}%
}(\overrightarrow{p})\epsilon_{m_{f}}(\overrightarrow{p})+\overrightarrow{p}%
.\overrightarrow{p})(1-\frac{|\overrightarrow{p}|^{2}}{(\epsilon_{m_{f}%
}(\overrightarrow{p})+m_{f})^{2}})-\nonumber\\
&  \text{ \ \ \ \ \ \ }2\frac{|\overrightarrow{p}|^{2}}{\epsilon_{m_{f}%
}(\overrightarrow{p})+m_{f}}{\Large )}b_{r}^{\dag}(\overrightarrow{p})\text{
}b_{r}(\overrightarrow{p})\frac{\epsilon_{m_{f}}(\overrightarrow{p})+m_{f}%
}{2\epsilon_{m_{f}}(\overrightarrow{p})}\nonumber\\
&  =\sum_{\overrightarrow{p},}\sum_{r=1,2}{\Large (}\varkappa\text{ }%
(\epsilon_{m_{f}}(\overrightarrow{p})\epsilon_{m_{f}}(\overrightarrow{p}%
)+\overrightarrow{p}.\overrightarrow{p})(\frac{|\overrightarrow{p}|^{2}%
+m_{f}^{2}-\overrightarrow{p}^{2}+2m_{f}\text{ }\epsilon_{m_{f}}%
(\overrightarrow{p})+m_{f}^{2}}{(\epsilon_{m_{f}}(\overrightarrow{p}%
)+m_{f})^{2}})-\nonumber\\
&  \text{ \ \ \ \ \ \ }2\frac{|\overrightarrow{p}|^{2}}{\epsilon_{m_{f}%
}(\overrightarrow{p})+m_{f}}{\Large )}b_{r}^{\dag}(\overrightarrow{p})\text{
}b_{r}(\overrightarrow{p})\frac{\epsilon_{m_{f}}(\overrightarrow{p})+m_{f}%
}{2\epsilon_{m_{f}}(\overrightarrow{p})}\nonumber\\
\text{ \ \ \ \ \ \ }  &  =\sum_{\overrightarrow{p},}\epsilon_{m_{f}%
}(\overrightarrow{p})\epsilon_{m_{f}}(\overrightarrow{p})b_{r}^{\dag
}(\overrightarrow{p})\text{ }b_{r}(\overrightarrow{p})\frac{1}{\epsilon
_{m_{f}}(\overrightarrow{p})}\nonumber\\
\text{ \ }  &  =\sum_{\overrightarrow{p},\text{ }r=1,2}\epsilon_{m_{f}%
}(\overrightarrow{p})b_{r}^{\dag}(\overrightarrow{p})\text{ }b_{r}%
(\overrightarrow{p})
\end{align}

Finally, for the negative energy massive term, a closely similar calculation
gives%
\begin{equation}
\int dx^{3}v_{m_{f}}^{\dagger}(x)\text{ }h(\partial)\text{ }v_{m_{f}}%
(x)=\sum_{\overrightarrow{p},\text{ }r=1,2}\epsilon_{m_{f}}(\overrightarrow{p}%
)\text{ }d_{r}^{\dag}(\overrightarrow{p})\text{ }d_{r}(\overrightarrow{p}).
\end{equation}

\end{document}